\documentclass[preprint,showpacs,titlepage,aps,prd,tightenlines,amsmath,byrevtex,nofootinbib]{revtex4}
\usepackage{graphicx}

\newcommand{\Dmq}{\Delta m^2}
\newcommand{\Sol}{\text{sol}}
\newcommand{\Atm}{\text{atm}}
\newcommand{\Sbl}{\textsc{sbl}}
\begin{document}
\baselineskip=17pt
\title{Signatures of sterile neutrino mixing in high-energy cosmic neutrino flux}

\author{Andrea Donini}
\email{E-mail: andrea.donini_at_uam.es}
\affiliation{Instituto F{\'i}sica Te{\'o}rica UAM/CSIC, Cantoblanco,
E-28049 Madrid, Spain}

\author{Osamu Yasuda}
\email{E-mail: yasuda_at_phys.metro-u.ac.jp}
\affiliation{Department of Physics, Tokyo Metropolitan University,
Hachioji, Tokyo 192-0397, Japan}

\begin{abstract}
Even after the negative result by the MiniBooNE experiment, we can
still have a (3+1)-scheme with one sterile neutrino whose mixing lies
within the allowed region by MiniBooNE.  In this note we discuss the
possible effects of such a scheme on the flavor ratio of high-energy
cosmic neutrinos from cosmologically distant astrophysical sources.
It is shown that in principle
there is still a chance to observe deviation from the
standard three flavor scenario in the flavor ratio of the high-energy
cosmic neutrinos.  It is proposed to see the energy spectrum to cope
with the theoretical uncertainties which were recently pointed out by
Lipari, Lusignoli and Meloni.  It is emphasized that the
$\nu_\mu-\nu_\tau$ ratio is relatively insensitive to the theoretical
uncertainties and therefore this ratio is the key ingredient to look
for the signatures of the sterile neutrino scheme.
Although the statistics of data from one source
in the next generation of neutrino telescopes are estimated
not to be sufficient to distinguish the three and four family schemes,
if we can gain statistics by, e.g., summing over data from many sources,
then it might be possible to
have a signature for the (3+1)-scheme.
\end{abstract}

\maketitle

\section{Introduction}
There has recently been much interest in high energy cosmic neutrinos from
cosmologically distant astrophysical sources, e.g., Active Galactic Nuclei (AGN) and Gamma Ray 
Bursts (GRB) (see, e.g., \cite{Halzen:2002pg}), motivated by their possible observation at
ongoing or future experiments such as ICECUBE \cite{ICECUBE}, BAIKAL \cite{BAIKAL}, NESTOR \cite{NESTOR},
ANTARES \cite{ANTARES}, NEMO \cite{NEMO}, RICE \cite{RICE}, AUGER \cite{AUGER}, EUSO \cite{EUSO}.
The flavor composition of the high energy cosmic neutrino flux has been extensively studied by many people
\cite{Learned:1994wg, Bento:1999bb, Athar:2000yw, Barenboim:2003jm, Beacom:2003nh, Beacom:2002vi, Beacom:2003eu, Dutta:2001sf, Keranen:2003xd, Awasthi:2007az, Xing:2007rz, Pakvasa:2007dc}, since the possibility of flavor tagging of high energy neutrinos was advanced in Ref. ~\cite{Learned:1994wg}.
A measurement of the flavor composition can, in principle, give important informations on the properties of the astrophysical neutrino source, on the leptonic mixing sector or point out new physics phenomena beyond the Standard Model 
of fundamental interactions. 
This claim, however, has been critically examined in Ref.~\cite{Lipari:2007su}, where it was shown that  it will be difficult  to 
learn something on the neutrino sources with the expected precision on high energy cosmic neutrino flavor 
measurements at ongoing and planned experiments. To illustrate the point, the authors of Ref.~\cite{Lipari:2007su}
consider in detail the case of neutrinos produced by $\gamma$ ray bursts (GRB)
and take into account their energy distribution according to Waxman and Bahcall \cite{Waxman:1997ti}
( in the framework of the "fireball model" of GRB, see for example Ref.~\cite{Fan:2008cg} and Refs. therein). 
The outcome of their analysis is that
 it is not clear if discussions on the astrophysical neutrino flavor composition make sense at all. 

We do not pretend in this paper to clarify if a measurement of the neutrino flavor fluxes is possible
in the absence of a precise knowledge of the source. We, more modestly, present an update of the analysis
that was performed in Ref.~\cite{Athar:2000yw} on the possibility to use such measurements at 
neutrino telescopes to constrain four neutrino models. 

Historically, four neutrino mass schemes became popular after the LSND group announced their data 
which suggest neutrino oscillation with the mass squared difference
$\Delta m^2\sim{\cal}O(1)$ eV$^2$ \cite{Athanassopoulos:1996jb,Athanassopoulos:1997pv,Aguilar:2001ty}.
The fourth neutrino has to be sterile, because the number of the light active neutrinos is three to be consistent
with the LEP data \cite{Yao:2006px}.  In Ref.~\cite{Athar:2000yw}, the neutrino telescope flavor data were used 
to constrain one specific four neutrino model, the so-called (2+2)-scheme, that could explain the LSND data
together with negative results at other experiments. Recently, the MiniBooNE experiment \cite{AguilarArevalo:2007it}
gave a negative result for neutrino oscillations with the mass squared difference 
$\Delta m^2\sim{\cal}O(1)$ eV$^2$ which was suggested by the LSND data.
While several scenarios have been proposed \cite{Barger:2005mh,
deGouvea:2006qd, Schwetz:2007cd, Nelson:2007yq} to account for both
the affirmative result by LSND and the negative one by MiniBooNE, four neutrino models are currently
unable to explain the LSND data.

However, sterile neutrino scenarios which satisfy all the experimental constraints except LSND are clearly possible.  
In Ref. \cite{Donini:2007yf}, a detailed analysis was performed on the allowed region of the mixing angles of the so-called (3+1)-scheme,  which satisfies all the constraints but LSND. In this paper we perform an analysis of the observed ratio
of the high energy neutrinos in the case of the (3+1)-scheme for the allowed region shown in Ref.~\cite{Donini:2007yf},
and discuss whether we can establish the signature of this scenario by looking at the ratios
$R_{e\mu} = (\nu_e+\bar{\nu}_e)/(\nu_\mu+\bar{\nu}_\mu)$ and $R_{\tau\mu} = (\nu_\tau+\bar{\nu}_\tau)/(\nu_\mu+\bar{\nu}_\mu)$ of the neutrino flux taking into account the uncertainties discussed in Ref.~\cite{Lipari:2007su}.

A similar analysis has been carried out in Ref.~\cite{Awasthi:2007az}. The main differences between 
this paper and Ref.~\cite{Awasthi:2007az} are the following: 
\begin{itemize}
\item 
We leave all the parameters of the  (3+1) four neutrino model (both angles and CP-violating phases)  free to vary, but only
within the presently allowed region, as from Ref.~\cite{Donini:2007yf}. We point out the critical role played by $\theta_{34}$
in the difference between three- and four-family mixing. However, since $\theta_{34}$ cannot exceed $\sim  35^\circ$, 
no striking sterile neutrino signal has to be foreseen at a neutrino telescope experiment. Searching sterile 
neutrinos at this generation of experiments will be a painful task, if possible at all.
\item 
We study the energy dependence of the neutrino flavor fluxes and of the flavor ratios in the case
of neutrinos from GRB, following the analysis of Ref.~\cite{Lipari:2007su}. We show that the uncertainties
induced by the neutrino flux energy spectrum in $R_{e\mu}$ makes impossible to use this observable 
to distinguish three- from four-family mixing. On the other hand, we find that $R_{\tau\mu}$ is much less
affected by the energy dependence. The flavor ratio $R_{\tau\mu}$ is
the optimal observable to look for sterile neutrinos. 
\item
We present analytic formul\ae \,  for the neutrino flavor fluxes and ratios expanded at second order
in the small parameters of the model, $\theta_{13}, \theta_{14}, \theta_{24}$ and $\eta = \pi/4 - \theta_{23}$. 
These formul\ae \,  help in the understanding of the numerical results presented in the paper. 
\end{itemize}

The paper is organized as follows.  In Sect.~\ref{sec:schemes} we briefly summarize the (3+1)-scheme 
and the allowed region of the mixing angles which are given by all the experimental data except
LSND.  In Sect.~\ref{sec:fluxes} we study the effects of the (3+1)-scheme on the flavor ratios 
$R_{e\mu}=(\nu_e+\bar{\nu}_e)/(\nu_\mu+\bar{\nu}_\mu)$ 
and $R_{\tau\mu}=(\nu_\tau+\bar{\nu}_\tau)/(\nu_\mu+\bar{\nu}_\mu)$. 
In Sect.~\ref{sec:energy} we discuss the energy dependence of the flavor ratios  in the
case of neutrinos produced in GRB's and examine whether deviations from the three-family expected values 
are large enough compared with the theoretical uncertainties of the neutrino flux given in Ref.~\cite{Lipari:2007su}.
In Sect.~\ref{sec:stats} we present a short discussion on the statistical requirements needed to 
identify a sterile neutrino signal using the flavor ratios.
In Sect.~\ref{sec:concl}, we eventually draw our conclusions. 
Analytic formul\ae \, at second order in  $\theta_{13}, \theta_{14}, \theta_{24}, \eta = \pi/4 - \theta_{23}$
and in the energy-dependent parameter $\epsilon (E_\nu)$ that is introduced to describe possible
deviations from ($\nu_e, \nu_\mu, \nu_\tau) = (1, 2 - \epsilon, 0)$
are shown in App.~\ref{app:formulae}.

\section{Four neutrino mass schemes}
\label{sec:schemes}
Four-neutrino schemes are classified into two classes: (3+1)- and
(2+2)- schemes, depending whether one or two mass state(s) are
separated from the other by a ${\cal O}(1)$ eV$^2$ mass-squared gap.
When one tries to account for the LSND data and for the negative results at
atmospheric and short baseline experiments with four-neutrino schemes, 
it has been found that we fall into problems irrespectively of which scheme we choose.

In (2+2)-schemes, the sterile neutrino contributes either to the solar or 
atmospheric oscillations. The fraction of sterile
neutrino contributions to solar and atmospheric oscillations
is given by $|U_{s1}|^2+|U_{s2}|^2$ and $|U_{s3}|^2+|U_{s4}|^2$, respectively,
where the mass squared differences $\Delta m^2_{21}$ and $|\Delta m^2_{43}|$
are assumed to be those of the solar and atmospheric oscillations.
The experimental results show that mixing among active neutrinos give dominant contributions
to both the solar and atmospheric oscillations (see, e.g., Ref. \cite{Maltoni:2004ei}).
In particular, in Fig.~19 of Ref.~\cite{Maltoni:2004ei} we can see that at the 99\% level
$\eta_s\equiv|U_{s1}|^2+|U_{s2}|^2 \le 0.25$ and $1-\eta_s=|U_{s3}|^2+|U_{s4}|^2\le 0.25$,
which contradicts the unitarity condition $\sum_{j=1}^4|U_{sj}|^2= 1$.
In fact the (2+2)-schemes are excluded at 5.1$\sigma$ CL \cite{Maltoni:2004ei}.
This conclusion is independent of whether we take the LSND data
into consideration or not.

In the (3+1)-scheme, on the other hand, in order to account for LSND without contradiction with
other disappearance experiments, the oscillation probabilities of the appearance and disappearance channels 
have to satisfy the following relation \cite{Okada:1996kw,Bilenky:1996rw}:
\begin{eqnarray}
\sin^22\theta_{\mbox{\rm\tiny LSND}}(\Delta m^2)
<\frac{1}{4}\,\sin^22\theta_{\mbox{\rm\scriptsize Bugey}}(\Delta m^2)
\cdot
\sin^22\theta_{\mbox{\rm\tiny CDHSW}}(\Delta m^2)
\label{relation31}
\end{eqnarray}
where $\theta_{\mbox{\rm\tiny LSND}}(\Delta m^2)$,
$\theta_{\mbox{\rm\tiny CDHSW}}(\Delta m^2)$,
$\theta_{\mbox{\rm\scriptsize Bugey}}(\Delta m^2)$ are the value of the two flavor
mixing angle as a function of the mass squared difference $\Delta m^2$ in the allowed region for LSND
($\bar{\nu}_\mu\rightarrow\bar{\nu}_e$), the CDHSW experiment \cite{Dydak:1983zq}
($\nu_\mu\rightarrow\nu_\mu$), and the Bugey experiment \cite{Declais:1994su}
($\bar{\nu}_e\rightarrow\bar{\nu}_e$).
From the early stages of the four neutrino schemes it has been known \cite{Okada:1996kw,Bilenky:1996rw}
that Eq. (\ref{relation31}) gives a strong constraint because there is very little region of the value of $\Delta m^2$
which satisfies Eq. (\ref{relation31}). While the significance to exclude the (3+1)-scheme changed
\cite{Barger:2000ch} when the LSND allowed region shifted slightly toward a smaller mixing region, difficulty to satisfy 
Eq. (\ref{relation31}) is basically the reason why this scheme is also disfavored.

A (3+2)-scheme with two sterile neutrino has also been proposed \cite{Sorel:2003hf} to account for
LSND, and it may be possible to reconcile the LSND and MiniBooNE data by introducing
a CP phase \cite{Karagiorgi:2006jf,Maltoni:2007zf}.
However, tension with the disappearance experiments such as CDHSW and Bugey always remains, 
as long as we take into account the LSND data.

In this paper we will consider a (3+1)-scheme without
taking the LSND data into account, e.g., we will assume a (3+1)-scheme
which satisfies all the negative constraints given by the appearance
experiments as it was done in Ref. \cite{Donini:2007yf}.
Then we no longer have the constraint (\ref{relation31})
and we have only the upper bound on the extra mixing angles,
as we will see below.
Apart from our strategy not to take the LSND constraint into account,
our framework is the same as the conventional (3+1)-scheme.

There has been discussions in which the mixing angles of four neutrino
schemes may be constrained by big-bang nucleosynthesis
\cite{Okada:1996kw,Bilenky:1998ne}, and if such arguments are applied,
then the mixing angles of sterile neutrinos would be very small, and
no significant signature would be expected in the flavor ratio of high
energy cosmic neutrinos.  However, it was shown in some model
\cite{Foot:1996qc} that neutrino oscillations themselves create large
lepton asymmetries which prevent sterile neutrinos from being in
thermal equilibrium, so it is not so clear whether the arguments in
\cite{Okada:1996kw,Bilenky:1998ne} hold.  At present, therefore, it is
fare to say that there is not yet general consensus on this issue (See
Ref. \cite{Cirelli:2004cz} and references therein.).
In this paper we will not impose cosmological constraints
on our scheme.

In four neutrino schemes, the $4\times4$ mixing matrix $U$
contains six mixing angles $\theta_{ij}$ and three Dirac CP phases
$\delta_i$ (where we ignore the three Majorana CP phases, 
since only Dirac CP phases can be measured in oscillation experiments).
There are plenty of different parametrizations of a general unitary $4\times4$ mixing matrix $U$. 
The following parametrization \cite{Maltoni:2007zf},
valid in the "atmospheric regime"
(i.e. where $ \Dmq_\Sol L/ E \to 0 $, $ \Dmq_\Atm L/ E \sim {\cal O}(1)$
and $|\Dmq_\Sbl L/E| \gg 1$), is particularly helpful to extract
the allowed region of the parameters of the present (3+1)-scheme
from experiments:
\begin{eqnarray}
    \label{eq:3+1param2}
    U =
    R_{34}(\theta_{34}) \; R_{24}(\theta_{24}) \;
    R_{23}(\theta_{23} ,\, \delta_3) \;
    R_{14}(\theta_{14}) \; R_{13}(\theta_{13} ,\, \delta_2) \; 
    R_{12}(\theta_{12} ,\, \delta_1) \,.
\label{rotation}
\end{eqnarray}
In Eq. (\ref{rotation})
$R_{jk}(\theta)\equiv\exp[\theta\Lambda_{jk}]$
and
$R_{jk}(\theta, \delta)\equiv\exp[-i(\delta/2)D_{jk}]
\exp[\theta\Lambda_{jk}]\exp[i(\delta/2)D_{jk}]$
are $4\times4$ rotation matrices, where
$i(\Lambda_{jk})_{\ell m}\equiv i(\delta_{j\ell}-\delta_{km})$
and
$(D_{jk})_{\ell m}\equiv\delta_{j\ell}\delta_{\ell m}
-\delta_{k\ell}\delta_{\ell m}$ are the generators
of the {\it su}(4) Lie algebra.
With this parametrization the phase $\delta_1$ drops in
the limit $\Delta m^2_{21}\to 0$, $U$
reduces to the mixing matrix in the standard three flavor
case when $\theta_{14}, \theta_{24}, \theta_{34} \to 0$
and all the phases disappear from the oscillation probabilities
in the one-mass dominance limit.

The mixing matrix elements in this parametrization are: 
\begin{eqnarray}
&
\left \{
\begin{array}{lll}
U_{e1} & = & c_{12} c_{13} c_{14} \\
U_{e2} & = & c_{13} c_{14} s_{12} e^{-i \delta_1} \\
U_{e3} & = & c_{14} s_{13} e^{-i \delta_2} \\
U_{e4} & = & s_{14}
\end{array}
\right. 
&
\\
&
\left \{
\begin{array}{lll}
U_{\mu 1} & = & - c_{23} c_{24} s_{12} e^{i \delta_1} 
- c_{12} \left [
c_{24} s_{13} s_{23} e^{i (\delta_2 - \delta_3)}  
+ c_{13} s_{14} s_{24} \right ] \\
U_{\mu 2} & = & c_{12} c_{23} c_{24} - s_{12} e^{- i \delta_1}
\left [ c_{24} s_{13} s_{23} e^{i (\delta_2 - \delta_3)}
+ c_{13} s_{14} s_{24}
\right ] \\
U_{\mu 3} & = & c_{13} c_{24} s_{23} e^{-i \delta_3}
 - s_{13} s_{14} s_{24} e^{-i \delta_2}\\
U_{\mu 4} & = & c_{14} s_{24}
\end{array}
\right .
&
\end{eqnarray}

\begin{eqnarray}
\left \{
\begin{array}{lll}
U_{\tau 1} & = & s_{12} e^{i \delta_1}
\left [ c_{34} s_{23} e^{i \delta_3} + c_{23} s_{24} s_{34} \right ]\\
&{\ }&-c_{12} \left \{ 
c_{13} c_{24} s_{14} s_{34} + s_{13} e^{i \delta_2} \left [
c_{23} c_{34} - s_{23} s_{24} s_{34} e^{-i \delta_3}
\right ]
\right \} \\
U_{\tau 2} & = & - c_{12} \left [ c_{34} s_{23} e^{i \delta_3}
 + c_{23} s_{24} s_{34} \right ]\\
&{\ }&- s_{12} e^{- i \delta_1} \left \{ 
c_{13} c_{24} s_{14} s_{34} + s_{13} e^{i \delta_2} \left [
c_{23} c_{34} - s_{23} s_{24} s_{34} e^{-i \delta_3}
\right ]
\right \} \\
U_{\tau 3} & = & -c_{24} s_{13} s_{14} s_{34} e^{-i \delta_2}
 + c_{13} \left [ 
c_{23} c_{34} - s_{23} s_{24} s_{34} e^{-i \delta_3}
\right ] \\
U_{\tau 4} & = & c_{14} c_{24} s_{34}
\end{array}
\right .
&&
\\
\left \{
\begin{array}{lll}
U_{s1} & = & s_{12} e^{i \delta_1}
\left [ c_{23} c_{34} s_{24} - s_{23} s_{34} e^{i \delta_3} \right ]\\
&{\ }&- c_{12} \left \{ 
c_{13} c_{24} c_{34} s_{14} - s_{13} e^{i \delta_2} \left [
c_{34} s_{23} s_{24} e^{- i \delta_3} + c_{23} s_{34}
\right ]
\right \} \\
U_{s2} & = & - c_{12} \left [ c_{23} c_{34} s_{24}
- s_{23} s_{34} e^{i \delta_3 }\right ]\\
&{\ }&- s_{12} e^{-i \delta_1} \left \{
c_{13} c_{24} c_{34} s_{14} - s_{13} e^{i \delta_2} \left [
c_{34} s_{23} s_{24} e^{- i \delta_3} + c_{23} s_{34}
\right ]
\right \} \\
U_{s3} & = & - c_{24} c_{34} s_{13} s_{14} e^{- i \delta_2} - c_{13}
\left [ 
c_{34} s_{23} s_{24} e^{- i \delta_3} + c_{23} s_{34}
\right ] \\
U_{s4} & = & c_{14} c_{24} c_{34}
\end{array}
\right .
&&
\end{eqnarray}
where $c_{ij} = \cos \theta_{ij}$ and $s_{ij} = \sin \theta_{ij}$.

The constraints from the affirmative data of the solar, KamLAND and
atmospheric neutrino experiments, and the
negative results of the short baseline experiments on
the (3+1)-scheme were examined in detail in Ref. \cite{Donini:2007yf}
and the allowed region for $\theta_{13}$, $\theta_{14}$,
$\theta_{24}$, $\theta_{34}$ are given in Fig.2 of
Ref. \cite{Donini:2007yf}.
From this figure we can see that $\theta_{13}$, $\theta_{14}$ and $\theta_{24}$ 
are constrained to be smaller than $\sim 10^\circ$, whereas  $\theta_{34}$ can be as large as $35^\circ$.
A fourth "small" parameter that will be useful for the analytic understanding of our results is
$\eta = \pi/4 - \theta_{23}$, constrained by atmospheric and LBL data to be $|\eta| \leq 10^\circ$.
In the analytic formulae we will consider $\theta_{13}, \theta_{14}$, $\theta_{24}$ and $\eta$ being of the same
order and expand in powers of the four.  

\section{the observed fluxes of each flavor}
\label{sec:fluxes}

The oscillation lengths of standard neutrino flavor transitions are
rather short with respect to the typical astrophysical distances. It
is, therefore, a good approximation to average over the oscillation
frequencies. The oscillation probabilities, thus, have an extremely
simple expression,
\begin{eqnarray}
P_{\alpha\beta} \equiv P(\nu_\alpha\to\nu_\beta)
= \sum_{i = 1}^N |U_{\alpha i}|^2 |U_{\beta i}|^2,
\label{prob}
\end{eqnarray}
where $N$ stands for the number of neutrino flavors.
It can be seen that $P_{\alpha\beta}$ does not depend on the neutrino
energy nor on the distance from the source. The neutrino flux of
flavor $\alpha$ at the detector is:
\begin{eqnarray}
\Phi_\alpha = \sum_{\beta= e,\mu,\tau} P_{\alpha\beta} \Phi^0_\beta \, ,
\end{eqnarray}
where $\Phi^0_\alpha$ is the neutrino flux of flavor $\alpha$ at the source.
Notice that, in the absence of a mechanism to produce sterile
neutrinos at the source, the sum runs only over active neutrinos.
However, in the four neutrino scheme discussed later,
we expect a sterile component at the detector
(albeit small
due to the severe bounds on the active-sterile mixing angles).

In the standard picture of neutrino creation from pion decays
after interactions between accelerated protons and photons,
the flavor composition at the source is expected to be approximately
$\Phi^0_e:\Phi^0_\mu:\Phi^0_\tau = 1:2:0$, as in the case
of the atmospheric neutrinos.
If this argument holds, then
the neutrino flavor fluxes at the detector are, therefore, 
\begin{eqnarray}
\Phi_\alpha = \sum_{i=1}^N |U_{\alpha i}|^2
\left [ |U_{e i}|^2 + 2 |U_{\mu i}|^2 \right ] \Phi_e^0 \, , 
\end{eqnarray}
with $\Phi_e^0$ the flux of electron neutrinos at the source.

\begin{figure}
\vspace*{32mm}
\hspace*{-12mm}
\includegraphics[scale=0.6]{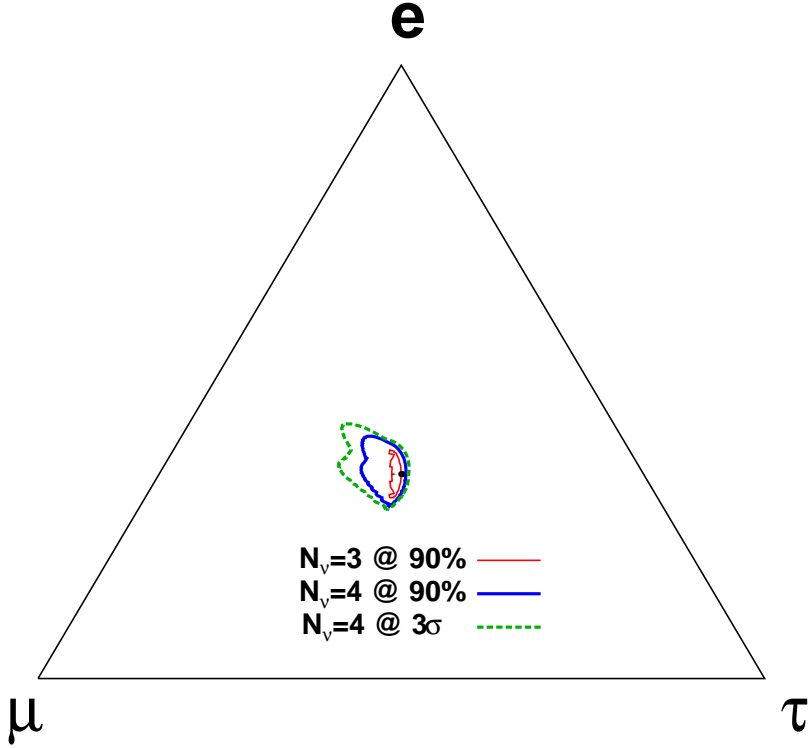}
\caption{\label{fig:triangle}
The allowed region of the flavor ratio $\Phi_e:\Phi_\mu:\Phi_\tau$ in the
triangle representation \cite{Athar:2000yw} for the three
flavor case (bounded by the red thin solid line) and the (3+1)-scheme (bounded
by the blue thick solid line) at 90\% CL.  The standard ratio of the
initial flux $\Phi^0_e:\Phi^0_\mu:\Phi^0_\tau = 1:2:0$ is assumed.
For the four flavor case
the region at 3$\sigma$ CL (bounded by the green dashed line) is also
depicted.  The black dot in the center indicates the case for
$\Phi_e:\Phi_\mu:\Phi_\tau=1:1:1$.}
\end{figure}

In the case of the standard three flavor mixing ($N=3$), when $\theta_{13}$ is small and
the atmospheric neutrino mixing angle is almost maximal, $|\theta_{23}-\pi/4|\ll1$, 
it is known \cite{Learned:1994wg} that the ratio of the flavor fluxes is approximately given by
\begin{eqnarray}
\Phi_e:\Phi_\mu:\Phi_\tau \simeq 1:1:1.
\end{eqnarray}
In the triangle representation of neutrino flavor fluxes, proposed in Ref.~\cite{Athar:2000yw}, 
the distance of a point from each of the sides of a regular  triangle represents the fraction of 
neutrinos of a given flavor, $\Phi_\alpha$, to the total neutrino flux, $\sum_\alpha \Phi_\alpha$. 
The allowed region in this representation for the three-family case is given by the red thin solid line  
in Fig.~\ref{fig:triangle}, where the allowed region at 90\%CL in Ref. \cite{Maltoni:2004ei} was used.

In the case of four neutrino schemes ($N=4$) the plot should in principle be three dimensional, 
because the flux of sterile neutrinos has to be also taken into consideration.  
As far as experiments are concerned, however, it is only the three active neutrinos that we can observe.
We normalize, thus,  the flux of each active neutrino by the total one of active neutrinos, so that we can still
use the triangle representation also in the four flavor case.

In 2000, when the (2+2)-scheme as well as all the solar neutrino solutions were still acceptable, the allowed regions in the triangle representation were relatively large (See Fig.~4 in Ref. \cite{Athar:2000yw}). One of the goal of this paper
is to update that analysis to the case of the (3+1)-scheme, when all experimental data but the LSND ones are
taken into account. We have performed both numerical and analytical studies to obtain the allowed region in 
the triangle representation for the (3+1)-scheme, when all the mixing angles $\theta_{13}$, $\theta_{14}$, $\theta_{24}$
and $\theta_{34}$ are left free to vary within the allowed region in Fig.~2 of Ref. \cite{Donini:2007yf}.

The allowed regions at 90\% CL (blue thick solid line) and at 3$\sigma$ CL (green dashed line)
for the (3+1)-scheme are shown in Fig.~\ref{fig:triangle}, where the numerical results have been obtained
assuming the standard ratio of initial neutrino fluxes, $\Phi^0_e:\Phi^0_\mu:\Phi^0_\tau = (1:2:0)$.
As it can be seen,  deviations of the (3+1)-scheme flavor fluxes from the three-family case are still noticeable\footnote{
In Ref.~\cite{Lipari:2007su}, it has been pointed out that for a neutrino flux produced by pion decay, the correct
flavor ratios are $\Phi^0_e:\Phi^0_\mu:\Phi^0_\tau = (1:1.86:0)$. We have numerically checked that the allowed regions
in this case are practically identical to those shown for the (1:2:0) case.}.

The results shown in Fig.~\ref{fig:triangle} are easily understood using the results from our analytical study.
At first order in the small parameters $\theta_{13}$, $\theta_{14}$, $\theta_{24}$ and $\eta\equiv\pi/4-\theta_{23}$, we get: 
\begin{eqnarray}
\left \{
\begin{array}{lll}
\Phi_e &=& \Phi_e^0 \left [ 1 +  \left ( \eta+ s_{13} \cos \varphi \cot 2 \theta_{12} \right )  \sin^2 2 \theta_{12}   \right ] \\
\Phi_\mu &=& \Phi_e^0 \left [  1 - \frac{1}{2}  \left ( \eta + s_{13} \cos \varphi \cot 2 \theta_{12} \right )  \sin^2 2 \theta_{12} \right ] \\
\Phi_\tau &=& c_{34}^2 \Phi_\mu \\
\Phi_s &=& s_{34}^2 \Phi_\mu
\end{array}
\right .
\label{flux-analytic}
\end{eqnarray}
where $\varphi = \delta_1 - \delta_2 + \delta_3$ is the only
combination of CP-violating phases present in this approximation.
Clearly, the sum of the four fluxes gives $\sum_\alpha \Phi_\alpha = 3 \Phi_e^0$, 
as it should be in the absence of neutrino decays. 
We can see that, at first order, deviations from (1:1:1:0) are governed by $\theta_{13}, \eta$ and $\theta_{34}$.
The only effect of active-sterile mixing arises, thus, through $\theta_{34}$, on which the experimental bounds
are the least severe ($\theta_{34}$ can be as large as $35^\circ$). A large $\theta_{34}$ can make a non trivial difference
between $\Phi_\mu$ and $\Phi_\tau$ that allows us to distinguish the (3+1)scheme from the three-family case.
If, on the other hand, $\theta_{34}$ were as small as $\theta_{14}$ and $\theta_{24}$, at leading order 
we would get a "three-family"-like result, $(\nu_e, \nu_\mu, \nu_\tau,  \nu_s) =
(1+O(\epsilon):1+O(\epsilon):1+O(\epsilon):O(\epsilon^2))$.
Eventually, in the limit $\theta_{34}\to 0$, Eq. (\ref{flux-analytic})
is reduced to the three flavor case with $\varphi\equiv\delta$.

Expressions for the flavor fluxes at second order in the small parameters $\theta_{13}, \theta_{14}, \theta_{24}$
and $\eta$ are shown in Appendix \ref{appendix1}. Here, we only point out that a dependence of the fluxes
on $\theta_{14}, \theta_{24}$ and CP-violating phases other than $\varphi$ arises at this order. 

\section{Energy dependence of the flavor fluxes}
\label{sec:energy}

Recently, the authors of Ref. ~\cite{Lipari:2007su} re-examined critically
the theoretical uncertainties in the predictions for the flavor ratio as well as 
its energy dependence. It has been shown that, quite generally, an  energy dependence
is induced in the neutrino flavor fluxes by the specific characteristics of the source. 
To illustrate their point, a detailed discussion of the energy dependence of neutrino flavor fluxes
originated from GRB is given.  We will follow in this section the approach of  Ref. ~\cite{Lipari:2007su} 
to determine the energy dependence of the flavor ratios in presence of sterile neutrinos. 

The analysis is performed following Waxman and Bahcall (see Ref.~\cite{Waxman:1997ti}) for neutrinos from GRB. 
The GRB itself is described in the so-called  "fireball model"  \cite{Fan:2008cg}.  Notice that an alternative theory has been proposed to explain $\gamma$-ray bursts, the so-called "cannonball model" \cite{Dar:2003vf}. 
In Ref.~\cite{Dar:2006dy}, however, it was shown that  no significant neutrino flux is to be expected from GRB's
in the framework of the cannonball model. 

Following Ref.~\cite{Waxman:1997ti}, the neutrino flux is supposed to mimic the photon flux\footnote{
We will not enter in the debate concerning the validity of this hypothesis, though, 
since we only use the neutrino spectrum obtained by Waxman and Bahcall to show how the
energy dependence of the neutrino flavor ratios makes extremely difficult to distinguish new physics effects
from standard three-family mixing.}  and to have a broken power-law spectrum. 
The neutrino flux depends on several phenomenological parameters that are needed to 
describe the (unknown) properties of the source:  
\noindent
(i) The proton spectrum $N_p(E_p)$ is assumed to depend on the proton energy $E_p$:
\begin{eqnarray}
N_p(E_p)\propto E_p^{-\alpha}.
\nonumber
\end{eqnarray}
(ii) The energy dependence of the number density $n_\gamma(\epsilon)$ of
the photons in the wind is described by the two exponents of the photon energy $\epsilon$:
\begin{eqnarray}
n_\gamma(\epsilon) \propto
\left \{
\begin{array}{ll}
 (\epsilon/\epsilon_{\rm b})^{-\beta_1}  
& {\rm for}~\epsilon \le \epsilon_{\rm b}, \cr
 (\epsilon/\epsilon_{\rm b})^{-\beta_2}  
& {\rm for}~\epsilon_{\rm b} < \epsilon  <
 \epsilon_{\rm max}, \cr
0
& {\rm for}~\epsilon \ge \epsilon_{\rm max},
\end{array}\right.
\nonumber
\end{eqnarray}
where $\epsilon_{\rm b}$ is the break energy and $\epsilon_{\rm max}$ is the cutoff energy.
\noindent
(iii) The energy loss of muons due to synchrotron radiation is characterized by
the magnetic field $B$ which is expressed in terms of the parameter $\epsilon_\mu$
\begin{eqnarray}
\epsilon_\mu=
\frac{E^\mu _{\rm syn}}{E^\ast}=8.4\times10^4 
\left(\frac{\rm Gauss}{B}\right)
\left( \frac{\epsilon_{\rm b}}{\rm KeV}\right),
\nonumber
\end{eqnarray}
where 
$E^\ast\simeq 6.9 \times 10^{13} (\epsilon_{\rm b}/{\rm KeV})^{-1}{\rm eV}$
is the proton threshold energy for inelastic interactions with photons
having the break energy $\epsilon_{\rm b}$, and $\epsilon_\mu=\infty$ corresponds 
to the case with no energy loss. For $\epsilon_b \sim 1$ KeV, we have $E^\ast = O(10^5)$ GeV.

When the parameters ($\alpha$, $\beta_1$, $\beta_2$) are varied
around the standard values ($\alpha=2$, $\beta_1=1$, $\beta_2=2$), the
ratio $(\Phi_e^0+\Phi_{\bar{e}}^0)/(\Phi_\mu^0+\Phi_{\bar{\mu}}^0)$
at the source becomes energy dependent, where $\Phi_{\bar{\alpha}}^0$
stands for the flux of antineutrino of flavor $\alpha$
at the source, and the ratio has significant deviation from 1/2 for lower
and higher values of the neutrino energy (cf. Fig.~14 in
Ref. \cite{Lipari:2007su}). 
When the muon energy loss due to synchrotron radiation is turned on,
an additional damping effect occurs and the ratio
$R_{e\mu}^0$ decreases at high neutrino energy (cf. Fig.~15 in Ref. \cite{Lipari:2007su}).

We have studied the impact of these effects on the flux ratio which is observed on the Earth in the case
of the three and four flavor mixing schemes. In general, if the flux ratio at the source is:
\begin{eqnarray}
(\Phi_e^0+\Phi_{\bar{e}}^0) : (\Phi_\mu^0+\Phi_{\bar{\mu}}^0) :
(\Phi_\tau^0+\Phi_{\bar{\tau}}^0)=1:\lambda(E_\nu):0,
\label{lambda}
\end{eqnarray}
where $\lambda$ is a function of the neutrino energy defined as
\begin{eqnarray}
\lambda (E_\nu) \equiv\frac{\Phi_\mu^0+\Phi_{\bar{\mu}}^0}{\Phi_e^0+\Phi_{\bar{e}}^0},
\nonumber
\end{eqnarray}
then the flavor ratios observed on the Earth can be written as
\begin{eqnarray}
R_{e\mu}&\equiv&\frac{\Phi_e+\Phi_{\bar{e}}}{\Phi_\mu+\Phi_{\bar{\mu}}}
=\frac{P_{ee}+\lambda P_{\mu e}}{P_{e\mu}+\lambda P_{\mu \mu}}
\label{ratioem}\\
R_{\tau\mu}&\equiv&\frac{\Phi_\tau+\Phi_{\bar{\tau}}}
{\Phi_\mu+\Phi_{\bar{\mu}}}
=\frac{P_{e\tau}+\lambda P_{\mu \tau}}
{P_{e\mu}+\lambda P_{\mu \mu}}
\label{ratiotm}
\end{eqnarray}
where we have used the properties $P_{\alpha \beta}=P_{\beta\alpha}$
and $P_{\bar{\alpha} \bar{\beta}}\equiv P(\bar{\nu}_\alpha\to\bar{\nu}_\beta)
=\sum_j|U_{\alpha j}^\ast|^2|U_{\beta j}^\ast|^2=P_{\alpha \beta}$ which can be 
easily derived from Eq. (\ref{prob}).

Although the flux ratios of $\bar{\nu}_e/\nu_e$
and $\bar{\nu}_\mu/\nu_\mu$ are in general different from unity,
the only necessary information to compute the flux
ratio observed on the Earth is the value of $\lambda$.
The value of $1/\lambda=R_{e\mu}^0$ is plotted in Figs.~14 and 15 in Ref. \cite{Lipari:2007su} 
as a function of the neutrino energy for the two cases mentioned above
(i.e. without and with the muon energy loss due to synchrotron radiation, respectively).

From those figures, we can see that the energy-dependence of the muon-to-electron ratio at the source
is rather small for most of the considered energy range. We can thus introduce a new parameter, $\epsilon (E_\nu)= 2 - \lambda (E_\nu)$, and expand in $\theta_{13}, \theta_{14}, \theta_{24}, \eta$ and $\epsilon$. 
At first order in these parameters, we get from eqs.~(\ref{ratioem}) and (\ref{ratiotm}): 
\begin{eqnarray}
\label{eq:ratioem1st}
R_{e\mu}^{\rm (4-fam)} &=& 1 + \frac{\epsilon}{2} \left ( 1 - \frac{3}{4} \sin^2 2 \theta_{12}  \right ) 
+ \frac{3}{2} \left ( \eta + \theta_{13} \cos \varphi \cot \theta_{12}  \right )  \sin^2 2 \theta_{12} 
= R_{e\mu}^{\rm (3-fam)} \\
\label{eq:ratiotm1st}
R_{\tau\mu}^{\rm (4-fam)} &=& c_{34}^2  = c^2_{34} R_{\tau\mu}^{\rm (3-fam)}
\end{eqnarray}
Expressions for $R_{e\mu}$ and $R_{\tau\mu}$ at second order in $\theta_{13}, \theta_{14}, \theta_{24}, \eta$ and $\epsilon$
can be found in App.~\ref{app:formulae}.

The relation between three- and four-family ratios  is immediately understood if we notice that, at first order in small parameters, we have the following simple relations between the mixing matrix elements
$|U_{\alpha j}|^2$ for four flavors and $|U_{\alpha j}^{(3)}|^2$ for three flavors:
\begin{eqnarray}
|U_{e j}|^2&=&\left\{
\begin{array}{l}
|U_{e j}^{(3)}|^2\quad (j=1,2,3)\cr
0\quad \quad \quad (j=4)
\end{array}
\right.
\label{3and411}\\
|U_{\mu j}|^2&=&\left\{
\begin{array}{l}
|U_{\mu j}^{(3)}|^2\quad (j=1,2,3)\cr
0\quad \quad \quad (j=4)
\end{array}
\right.
\label{3and412}\\
|U_{\tau j}|^2&=&\left\{
\begin{array}{l}
c^2_{34}|U_{\tau j}^{(3)}|^2\quad (j=1,2,3)\cr
s^2_{34}\qquad \quad \quad (j=4)
\end{array}
\right.
\label{3and413}\\
|U_{s j}|^2&=&\left\{
\begin{array}{l}
s^2_{34}|U_{\tau j}^{(3)}|^2\quad (j=1,2,3)\cr
c^2_{34}\qquad \quad \quad (j=4)
\end{array},
\right.
\label{3and414}
\end{eqnarray}
where the combination of the CP phases
$\varphi = \delta_1-\delta_2+\delta_3$ becomes $\delta$ 
in the standard parametrization in the three flavor case.

\begin{figure}
\vspace*{-10mm}
\includegraphics[scale=0.45]{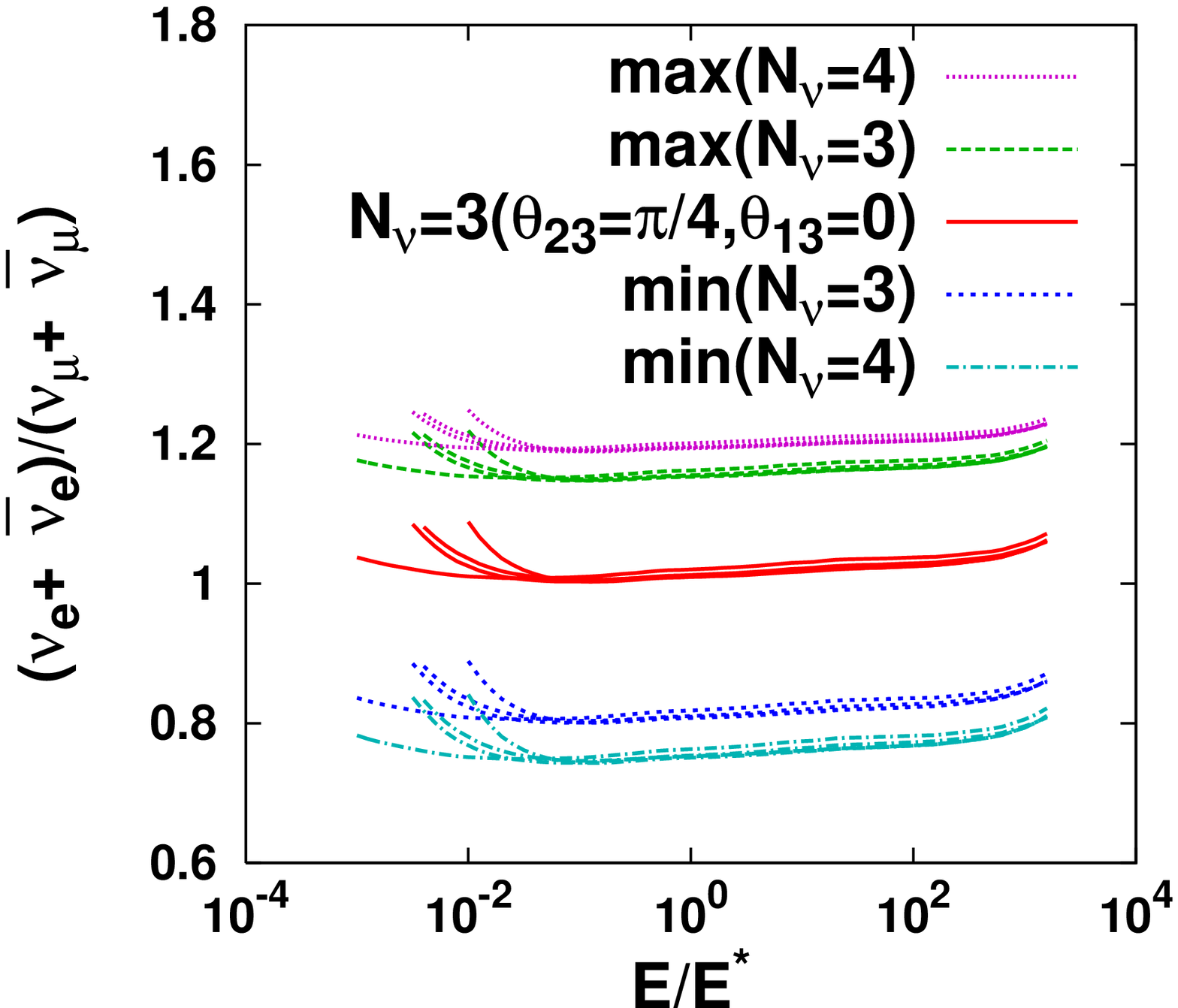}
\includegraphics[scale=0.45]{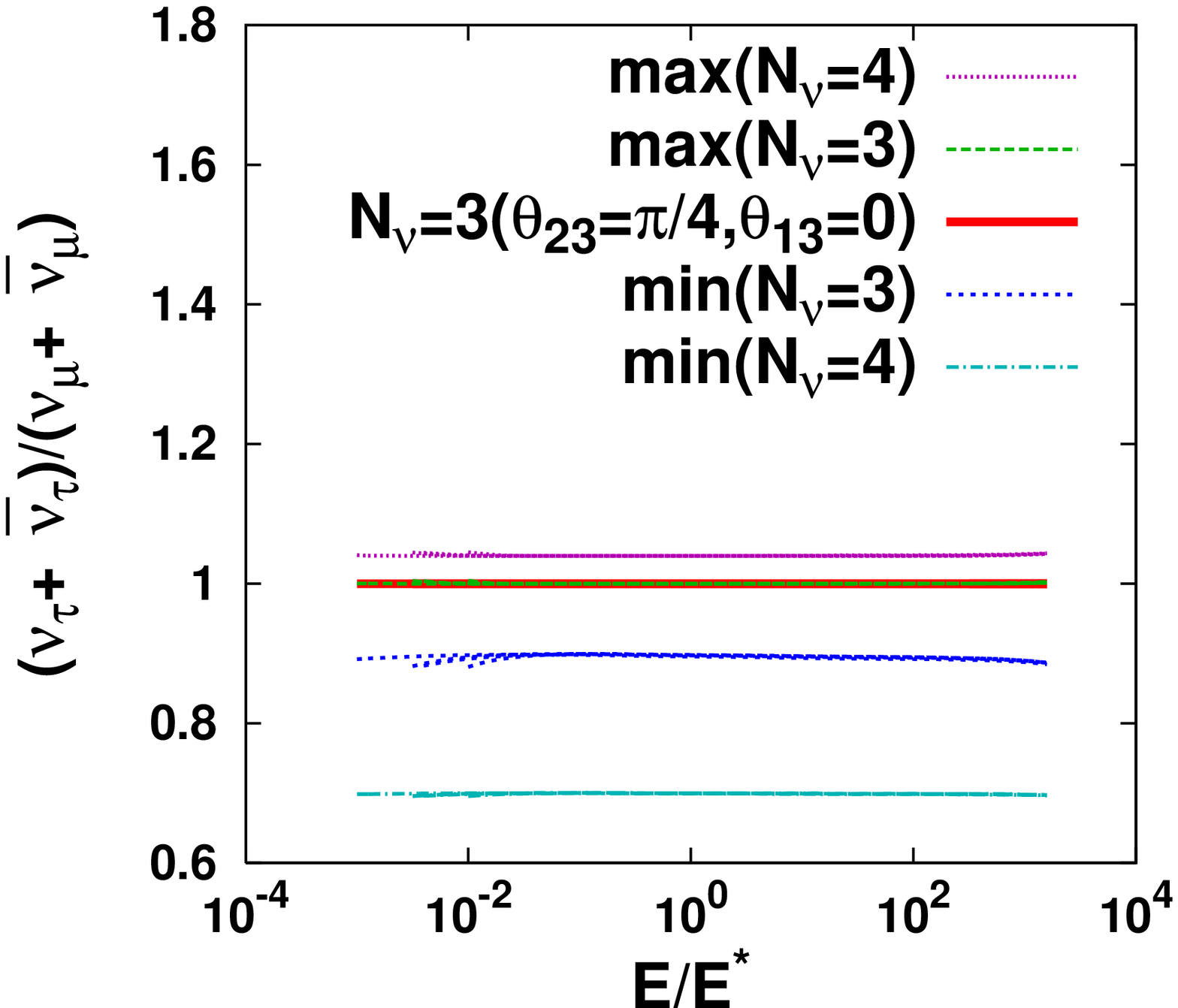}
\vspace*{5mm}
\caption{\label{fig:fig14}
The observed ratios $(\nu_e+\bar{\nu}_e)/(\nu_\mu+\bar{\nu}_\mu)$ and $(\nu_\tau+\bar{\nu}_\tau)/(\nu_\mu+\bar{\nu}_\mu)$ for neutrinos from a GRB source as functions of the neutrino energy.  The five cases (the three flavor case with the best-fit oscillation parameters; the three flavor cases which give the maximum and minimum ratios; the four flavor cases which give the maximum and minimum ratios) are considered with the four reference parameters $(\alpha=2, \beta_1=1, \beta_2=2)$, 
$(\alpha=2, \beta_1=0.65, \beta_2=2)$, $(\alpha=2, \beta_1=1, \beta_2=2.4)$, $(\alpha=2.4, \beta_1=1, \beta_2=2)$ in
Fig.~ 14 of Ref. \cite{Lipari:2007su} for each case.}
\end{figure}

From eqs.~(\ref{eq:ratioem1st}) and (\ref{eq:ratiotm1st}) we see that an energy dependence in the $\nu_e/\nu_\mu$ ratio 
arises  already at first order in small parameters, whereas the $\nu_\tau/\nu_\mu$ ratio is less sensitive to the energy
dependence of the source. On the other hand, deviations from the three-family scenario should be bigger in the
latter ratio than in the former, as a consequence of the less stringent bounds that we have on $\theta_{34}$ with 
respect to $\theta_{14}$ and $\theta_{24}$. This can be clearly seen in Fig.~\ref{fig:fig14}, where we plot  the maximal 
and minimal values assumed by $R_{e\mu}$ (left) and $R_{\tau\mu}$ (right) as a function of $E/E^\ast$  in the absence 
of the muon energy loss due to synchrotron radiation, when varying the three-family mixing angles ($\theta_{13}, \theta_{23}$
and $\theta_{12}$), the active-sterile mixing angles ($\theta_{14}, \theta_{24}$ and $\theta_{34}$) and
the CP-violating phases $\delta_1, \delta_2$ and $ \delta_3$ within the presently allowed region 
(see Ref. \cite{Maltoni:2004ei}). Notice that, when the parameters ($\alpha, \beta_1, \beta_2$) are varied,
these uncertainties give a little energy dependence to the observed ratio $R_{e\mu}$ but not to $R_{\tau\mu}$,
as it was expected from the previous discussion. The difference between the three- and four-family mixing
is not large in $R_{e\mu}$, but it can be remarkable in $R_{\tau\mu}$.

\begin{figure}
\includegraphics[scale=0.45]{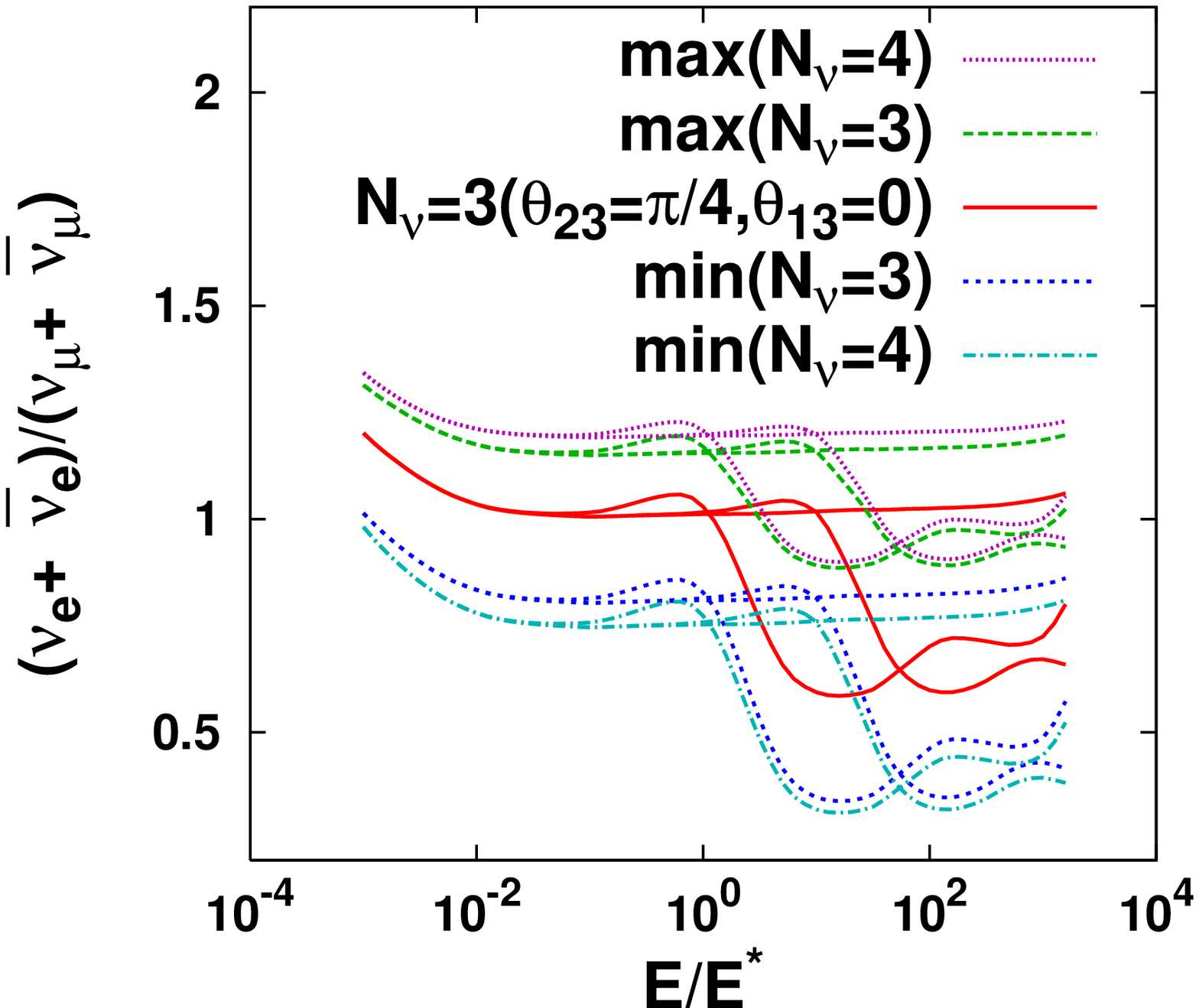}
\includegraphics[scale=0.45]{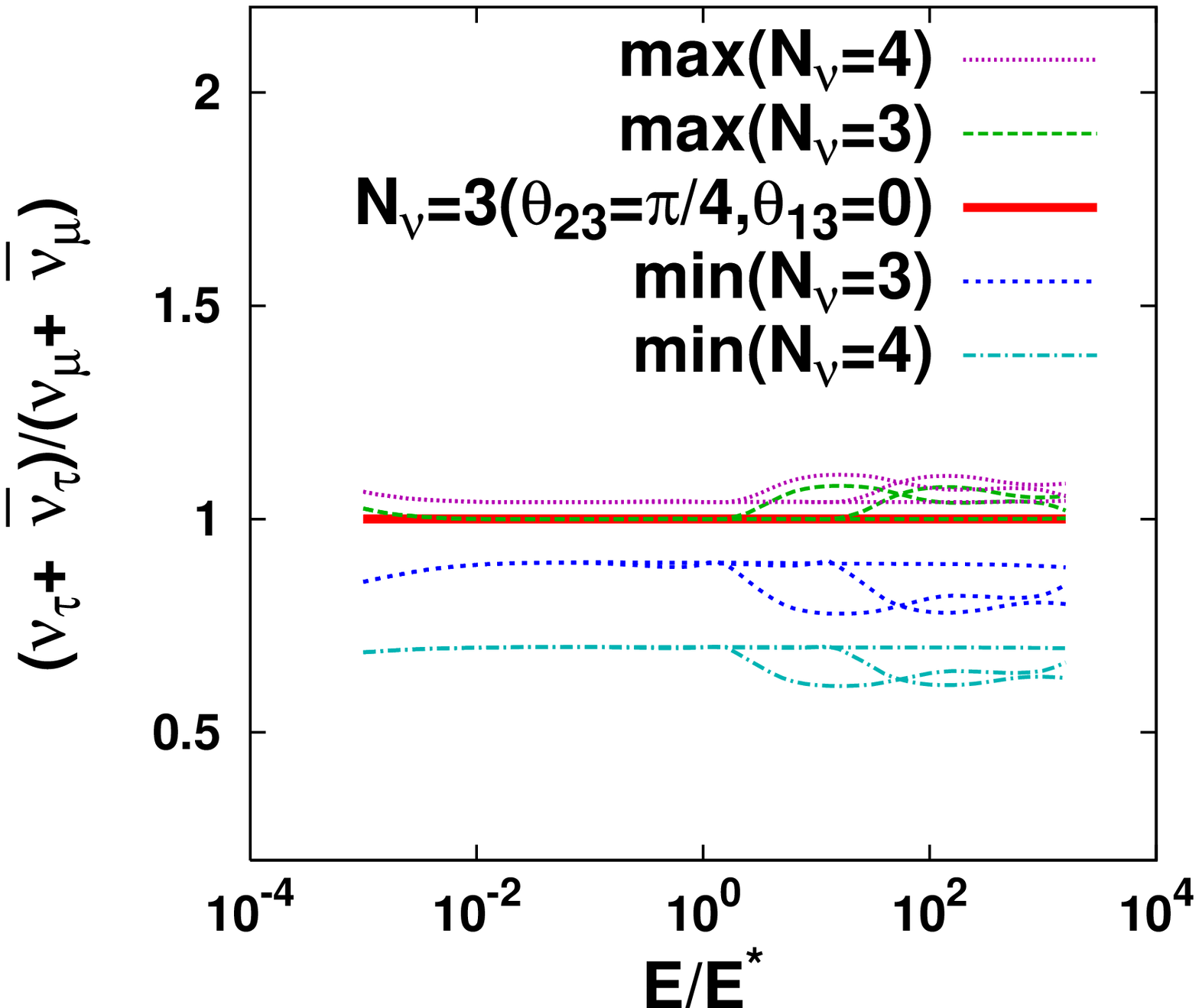}
\vspace*{5mm}
\caption{\label{fig:fig15}
The same as Fig.~\ref{fig:fig14} when the effect
of the synchrotron losses of high energy muons are taken into account.
The five cases are
considered with the reference parameters $\epsilon_\mu=\infty,
30, 3$ in Fig.~15 of Ref. ~\cite{Lipari:2007su} for each case.
}
\end{figure}

In Fig.~\ref{fig:fig15} we present the maximal and minimal values assumed by $R_{e\mu}$ (left) and $R_{\tau\mu}$ 
(right) as a function of $E/E^\ast$ in presence of the muon energy loss due to synchrotron radiation 
(taken from Fig.~15 of Ref.~\cite{Lipari:2007su}). 
In Fig.~\ref{fig:fig15}(left) we can see that, in presence of muon damping effects, the energy dependence 
of $R_{e\mu}$ becomes too large to disentangle new physics signals from the underlying uncertainties on the three-family mixing matrix parameters. Notice that theoretical uncertainties which are not taken into account in Ref. \cite{Lipari:2007su} can only worsen this result, making extremely difficult (if not impossible) to use $R_{e\mu}$ to look for new physics. 
On the other hand, in Fig.~\ref{fig:fig15}()right) we can see that the energy
dependence induced in $R_{\tau\mu}$ by the muon energy loss is much milder than for  $R_{e\mu}$. 
This flavor ratio could still be used to distinguish three- from four-family mixing when $\theta_{34}$ is near to its 
allowed upper bound. This is a further indication that $R_{\tau\mu}$ is the relevant observable
to look for a signature of the (3+1)-scheme.

Notice, however, that the energy domain for which muon damping effects are negligible is for $E < E^\ast$. 
For $E^\ast \sim 10^5$ GeV (taking $\epsilon_b = 1$), the energy range  for which it could be easier to use
 this flavor ratio to look for new physics is below the threshold for $\tau$ production ($E_\tau > 10^6$ GeV, 
 see Refs.~\cite{Learned:1994wg,Beacom:2003nh}).  It should be stressed, on the other hand, that even
 for $E > E^\ast$ the three- and four-family $R_{\tau\mu}$ ratios do not overlap (for maximal $\theta_{34}$)
 and therefore, with statistics large enough, the two models could be distinguished\footnote{This is not true for
 the $R_{e\mu}$ ratio.}. This is the main result of this paper. 

\section{Flavor tagging and statistical requirements}
\label{sec:stats}

It is now mandatory to discuss the statistical requirements needed to 
take advantage of high-energy $\nu$ from astrophysical sources to look 
for signals of sterile neutrinos. 

First of all, we remind how different flavor could be identified in a neutrino telescope experiment. 
As it was shown in Ref.~\cite{Beacom:2003nh}, muon-like events can be identified
with a good efficiency by looking for a single up-going muon entering the detector, 
whose energy can be estimated from the amount of \v Cerenkov light produced by photons and $e^\pm$ pairs
radiated from the muon track.  These are called "track events". 

A second class of events are those in which the release of  a large amount of \v Cerenkov light, generated
by a shower induced by a neutrino interaction in the detector volume, is observed. 
This signal is produced mainly by $\nu_e (\bar \nu_e)$ CC interactions, albeit with a (significant) contribution 
by $\nu_\mu$ and $\nu_\tau$ CC interactions and NC interactions of the three active neutrino flavors. 
A good knowledge of  "track events" and "shower events" is needed to have a handle on the $R_{e\mu}$
flavor ratio (see, for example, Ref.~\cite{Pakvasa:2007dc}).
Moreover, the $\nu_e$ and $\nu_\mu$ fluxes must be known with a good precision to quantify the high-energy atmospheric
neutrino background\footnote{Notice that $7 \times 10^5$ atmospheric neutrino
events with energies $E_\nu \geq 100$ GeV are expected in 10 years of data taking at IceCube, \cite{GonzalezGarcia:2005xw}.}  that is relevant for $E_\nu \leq 10^5$ GeV.  However, as it can be seen in Figs.~\ref{fig:fig14}(left) 
and \ref{fig:fig15}(left), the ratio of "shower" to "track events" is strongly dependent on the spectral shape of the flavor fluxes. 
This introduces a further problem in the utilization of this observable to look for new physics beyond the three-family 
neutrino mixing scenario or to pin down the details of the astrophysical source of the high-energy neutrinos.

Eventually, $\nu_\tau$ CC interactions can be singled out by looking for "double bang" or "lollipop" events 
\cite{Learned:1994wg} . The former are events in which two distinct energy releases are observed inside the 
detector volume, with space and time separation consistent with the propagation and decay of an energetic $\tau$. 
The latter consist of events in which the second "double bang" shower is observed along the $\tau$ track, 
but the first is missed. Both event classes are specific to $\nu_\tau$ CC interactions. Inverted lollipops (where the second shower is missed and the first one is observed) are not useful to identify $\tau$ events, since they can be confused with an energetic $\nu_\mu$ CC interaction in which a hadronic shower is reconstructed together with a muon track. 
The ratio of the sum of "double bang" and "lollipop events" to "track events" is, thus, a measure of $R_{\tau\mu}$.

We should now try to specify the statistical requirements needed to use flavor ratios at neutrino telescopes
(measurable as explained above) to look for new physics from astrophysical high-energy neutrinos.

Two sources have been considered in this paper: AGN's and GRB's. For both sources, we lack of an accurate model
to predict the expected neutrino rates, and we can only compute their flavor ratios. This is useless, however, if we cannot
establish if the observation of a deviation of flavor ratios from the three-family neutrino mixing expectation can indeed be 
a signal of new physics or it can just be ascribed to statistical fluctuations of the standard values. 

In Ref.~\cite{Lipari:2006uw}, the present catalogues of TeV $\gamma$-ray sources were used 
to infer the expected neutrino fluxes to be measured at neutrino telescope experiments. 
Indeed, the hadronic interactions that are the source of astrophysical neutrino fluxes also create a large number
of $\pi^0$ and $\eta$ particles. These particles decay eventually to photons, generating a high-energy $\gamma$ flux.
For a transparent source in which photons are produced mainly by this process,  the relation that can be inferred on the neutrino flux by the observation of a high-energy photon flux is robust\footnote{It must be reminded, however, that photons can also be produced through inverse Compton scattering of relativistic electrons on radiation fields.  In this case, the hadronic component is poorly measured and the neutrino flux can be much smaller than the photon flux. Eventually, for an opaque source, the photons can be absorbed inside the medium and subsequently emerge again at lower frequencies.}. 
The neutrino flux is as large as the photon flux, at least.  

The three brightest $\gamma$-ray sources detected by the HESS telescope 
are the Crab nebula, RXJ1713.7-3946 and Vela Junior \cite{Aharonian:2005jn,Aharonian:2005kn}.
Their integrated flux above 1 TeV is $(2.1,2.0,1.9) \times 10^{-11}$ (cm$^2$ s)$^{-1}$, respectively.
The photon emission of the last two objects is believed to be of hadronic origin. Considering a
photon flux half that of RXJ1713.7-3946 and Vela Junior, produced through hadronic mechanism, the
corresponding neutrino flux above 1 TeV  is $\Phi_\nu (> 1 \, {\rm TeV}) \sim 10^{-11}$ (cm$^2$ s)$^{-1}$. 
Using this neutrino flux, with an unbroken  power law spectrum, the total number of $e, \mu$ and $\tau$ contained
events that can be observed in a km$^3$ water equivalent detector is 10.3, 9.6 and 2.9 per year, respectively.
The expected flux of up-going $\nu$-induced muons is $\Phi_\mu = 5.6$ (km$^2$ year) $^{-1}$. 
These numbers take into account the different neutrino flavor energy spectra, as in Fig.~3 of Ref.~\cite{Lipari:2006uw}, 
the neutrino-nucleon cross-sections and the Earth absorption, but not the detector efficiency to identify a given
neutrino flavor. Similar results have been found in Refs.~\cite{Kappes:2006fg, Kistler:2006hp}.

The photon emission of the brightest AGN's present in the TeV $\gamma$-ray catalogue (Mkn421 and Mkn501)
\cite{EGRET} is  strongly varying with time. In the emission peaks, their average photon flux above 1 TeV can be several times  that of the Crab nebula.  The extrapolation from the photon flux to the neutrino flux is difficult, however, since we have no indication on the hadronic or leptonic origin of the photon emission.

Considering for AGN's and GRB's a flux similar to that of a galactic point source of the example given 
in Ref.~\cite{Lipari:2006uw}, we should get $O(100)$ $\nu_e$ and $\nu_\mu$ events and $O(30)$ $\nu_\tau$ events
after 10 years of data taking at a km$^3$ detector. Similar statistics has been used in Ref.~\cite{Maltoni:2008jr} to 
study neutrino decay at neutrino telescopes. It is easy to see that such statistics is too small to disentangle 
a (3+1) sterile neutrino effect from the statistical fluctuations of the three-family neutrino signals, even in the
case of $\theta_{34}$ near to its experimental limit, $\theta_{34} \sim 35^\circ$. 
The uncertainties on the three-family parameters within the presently allowed region in $R_{\tau\mu}$ are $O(10\%)$. 
The (gaussian) statistical fluctuations in $R_{\tau\mu}$ (after 10 years of data taking in the case of an ideal detector,
given the numbers above) can be as large as 20\% at 1$\sigma$. 
The expected three-family $R_{\tau\mu}$ flavor ratio is, thus, 
$R_{\tau \mu}^{\rm (3-fam)} \simeq 0.30 \pm 0.03 ({\rm theo}) \pm 0.06 ({\rm stat})$, to be compared with 
$R_{\tau\mu}^{\rm (4-fam)} (\theta_{34} = 35^\circ) \simeq c^2_{34} R_{\tau \mu}^{\rm (3-fam)} \sim 0.2$.
This means that it is extremely difficult that (3+1) sterile neutrinos produced by a single source
can be observed and unequivocally distinguished from theoretical and statistical fluctuations of the three-family mixing scenario in the next generation of neutrino telescopes.  A factor 30 more statistics is needed to detect unambiguously
a sterile neutrino signal with maximal $\theta_{34}$. A possible way out of this problem is to integrate over 
all the galactic and extragalactic sources (or, for example, to sum over many GRB's events of similar intensities)\footnote{The authors thank P. Lipari for discussions which lead to these ideas.}.
In this case, it is possible that the next generation of neutrino telescopes could observe or constrain 
the four-family mixing model by looking at the $R_{\tau\mu}$ flavor ratio. No signal can be expected from the 
$R_{e\mu}$ one.

\section{Conclusions}
\label{sec:concl}

We have studied the effects of the (3+1)-scheme, which is constrained
by all the experimental data except LSND, on the flux ratio of the high
energy cosmic neutrinos.  In principle,
if a large number of events are accumulated
in the future, there is still a chance to have a signature of
the (3+1)-scheme.
Furthermore, we have taken into account the theoretical
uncertainties which were discussed in Ref. \cite{Lipari:2007su}
in the case of high energy neutrinos from GRB.
While these uncertainties make the prediction for
the observed ratio $R_{e\mu}=(\nu_e+\bar{\nu}_e)/(\nu_\mu+\bar{\nu}_\mu)$
ambiguous, the ratio
$R_{\tau\mu}=(\nu_\tau+\bar{\nu}_\tau)/(\nu_\mu+\bar{\nu}_\mu)$
is less sensitive to the uncertainties and is sensitive to
the sterile neutrino mixing angle $\theta_{34}$.
The energy spectrum of these flux ratios can also
give us the information on the theoretical uncertainties,
particularly that of the muon energy loss.
While the statistical errors of data from a galactic point source
in the next generation of neutrino telescopes are estimated
to be too large to distinguish the three and four family schemes,
if we can gain statistics by, e.g., summing over data from many sources,
then it might be possible to
have a signature for the (3+1)-scheme in
observations of the high energy cosmic neutrinos.
More detailed studies on the systematic errors
to identify $\nu_\tau$ as well as the theoretical
uncertainties would be necessary to derive quantitative conclusions.

\appendix
\section{analytic expressions for the flavor fluxes and flavor ratios at second order\label{appendix1}}
\label{app:formulae}

At second order in $\theta_{13}$,
$\theta_{14}$, $\theta_{24}$ and $\eta\equiv\pi/4-\theta_{23}$,
the picture is far more entangled. Both $\theta_{14}$
and $\theta_{24}$ are present, together with the three CP-violating
phases. Notice that $s_{13}$ always appears in combination with the phase factor $\cos \varphi$, 
both in the expressions for the flavor fluxes and in the flavor ratios.We therefore define 
the reduced parameter  $\bar s_{13} = s_{13} \cos \varphi$, where $\varphi = \delta_1 - \delta_2 + \delta_3$
as in eq.~(\ref{flux-analytic}).
However, at second order, also the phases $\delta_1$ and $\delta_1 + \delta_3$ appear, and thus
three independent CP-violating phases are present at this order in the flavor fluxes expressions.

Notice that our results reduced to the corresponding three-family expressions at second order in $\theta_{13}$ 
and $\eta$ presented in Ref.~\cite{Pakvasa:2007dc} for $\theta_{14} = \theta_{24} = \theta_{34} = 0$.

For the electron neutrino flux at the detector, we get:
\begin{eqnarray}
\Phi_e/\Phi^0 &=& \left \{ 
1 + ( \eta + \bar s_{13} \cot 2 \theta_{12} )  \sin^2 2 \theta_{12}  \right . \nonumber \\
&-& \left . 2  s_{14}^2  +  \frac{1}{2}  \left (  s_{14}^2 - s_{24}^2 \right ) \sin^2 2 \theta_{12}
+ \frac{\sqrt{2}}{2} \sin 4 \theta_{12} s_{14} s_{24} \cos \delta_1
\right \} 
\end{eqnarray}

For the muon neutrino flux at the detector, we have: 
\begin{eqnarray}
\Phi_\mu/\Phi^0 &=& \left \{ 
1 - \frac{1}{2} \left ( \eta + \bar s_{13} \cot 2 \theta_{12} \right ) \sin^2 2 \theta_{12}
\right . \nonumber \\
&+& \bar s_{13}^2 \sin^2 2 \theta_{12}  - \eta \bar s_{13} \sin 4 \theta_{12}
+  \eta^2 \left ( 4 -  \sin^2 2 \theta_{12} \right ) 
\nonumber \\
&-& \left . 2 s_{24}^2 
-  \frac{1}{4}  (s_{14}^2 - s_{24}^2) \sin^2 2 \theta_{12}  
-\frac{\sqrt{2}}{4}   \sin 4 \theta_{12} s_{14} s_{24} \cos \delta_1 
\right \}
\end{eqnarray}

For the tau neutrino flux at the detector, we have: 
\begin{eqnarray}
\Phi_\tau/\Phi^0 &=& 
c_{34}^2 \left \{ 
1 - \frac{1}{2}  \left ( \eta  +    \bar s_{13}  \cot 2 \theta_{12} \right )  \sin^2 2 \theta_{12} 
\right . \nonumber \\
&-& \bar  s_{13}^2 \sin^2 2 \theta_{12} 
+\eta \bar s_{13} \sin 4 \theta_{12}
-   \eta^2 (4 - \sin^2 2 \theta_{12})  
 \nonumber \\
&-&\left .   s_{24}^2 -  \frac{1}{4} (s_{14}^2 - s_{24}^2) \sin^2 2 \theta_{12}
 - \frac{\sqrt{2}}{4} \sin 4 \theta_{12} s_{14} s_{24} \cos \delta_1 
\right \}
\nonumber \\
&+& s_{34}^2 \left ( 2 s_{14}^2 + 3 s_{24}^2 \right ) \nonumber \\
&+& (c_{34} s_{34} ) \left \{
s_{24} \cos \delta_3 \left [ \eta
\left ( 4 - \sin^2 2 \theta_{12} \right )
- \frac{1}{2} \bar s_{13} \sin 4 \theta_{12}  \right ] 
\right . \nonumber \\
&+& \left . \sqrt{2} s_{14} \cos (\delta_1 + \delta_3) \sin^2 2 \theta_{12} 
\left ( \eta \cot 2 \theta_{12} - \bar s_{13} ) \right ]
\right \}
\end{eqnarray}

Eventually, for the sterile neutrino flux at the detector, we get: 
\begin{equation}
\Phi_s = \Phi_\tau (\theta_{34} \to \theta_{34} + \pi/2)
\end{equation}

Flavor ratios at first order in  $\theta_{13}$, $\theta_{14}$,  $\theta_{24}$, $\eta\equiv\pi/4-\theta_{23}$
and $\epsilon(E_\nu) = 2 - \lambda(E_\nu)$ have been presented in eqs.~(\ref{eq:ratioem1st})-(\ref{eq:ratiotm1st}). 
At second order in the same quantities we get:

\begin{eqnarray}
\label{eq:ratioem2nd}
R_{e\mu}^{\rm (4-fam)} &=& 1 + \frac{\epsilon}{2} \left ( 1 - \frac{3}{4} \sin^2 2 \theta_{12}  \right ) 
+ \frac{3}{2} ( \eta  + \bar \theta_{13} \cot 2 \theta_{12}  ) \sin^2 2 \theta_{12} 
\nonumber \\
&-& 2 (\theta_{14}^2 + \theta_{24}^2) \left ( 1 - \frac{3}{8} \sin^2 2 \theta_{12} \right ) 
+ \frac{3}{4} \sqrt{2} \sin 4 \theta_{12} \theta_{14} \theta_{24} \cos \delta_1 
\nonumber \\
&-& \frac{1}{4} \bar \theta_{13}^2  \sin^2 2 \theta_{12} (1 + 3 \sin^2 2 \theta_{12})
+ \eta \bar \theta_{13} \left ( 1 + \frac{3}{4} \sin^2 2 \theta_{12} \right ) \sin 4 \theta_{12}
     \nonumber \\
&-& 4 \eta^2 \left ( 1 -  \frac{1}{4} \sin^2  \theta_{12} - \frac{3}{16} \sin^4 2 \theta_{12} \right ) 
\nonumber \\
&+& \frac{3}{32} \epsilon \left ( \eta +  \bar \theta_{13} \tan 2 \theta_{12} \right ) \sin^2 4 \theta_{12}
\end{eqnarray}
and
\begin{eqnarray}
\label{eq:ratiotm2nd}
R_{\tau\mu}^{\rm (4-fam)} &=& c_{34}^2
                \left \{ 1 - 8 \eta^2 \left ( 1 - \frac{1}{4} \sin^2 2 \theta_{12} \right ) 
                 + \frac{1}{2} ( \epsilon \eta - 4 \bar \theta_{13}^2 ) \sin^2 2 \theta_{12}
                \right .
\nonumber \\
&+& \left .  \frac{1}{4} ( \epsilon + 8 \eta  ) \bar \theta_{13}  \sin 4 \theta_{12}  \right \}
\nonumber \\
&+& \sin 2 \theta_{34} \left \{ \frac{1}{\sqrt{2}} \theta_{14} \cos (\delta_1 + \delta_3)
\left ( \eta \cos 2 \theta_{12} -  \bar \theta_{13}  \sin 2 \theta_{12} \right ) \sin 2 \theta_{12} \right .
\nonumber \\
&+& \left . \theta_{24} \cos \delta_3 \left [
2 \eta \left ( 1 - \frac{1}{4} \sin^2 2 \theta_{12} \right ) - \frac{1}{8 }\epsilon \sin^2 2 \theta_{12} 
- \frac{1}{4} \bar \theta_{13}  \sin 4 \theta_{12}    \right ] \right \} 
\nonumber \\
&+& \left \{ \theta_{24}^2 + 2 s_{34}^2 (\theta_{14}^2 + \theta_{24}^2) \right \}
\end{eqnarray}

\section*{Acknowledgments}
We acknowledge useful discussions with A. De Rujula, P. Lipari and M. Lusignoli.
O.Y. would like to thank the Instituto de F\'{\i}sica Te\'orica UAM/CSIC  for the hospitality during part of this work.
This research was supported in part by the JSPS Bilateral Joint Projects (Japan-Spain) and a Grant-in-Aid for Scientific
Research of the Ministry of Education, Science and Culture, \#19340062. A.D. acknowledges funding from the 
Consejo Superior de Investigaciones Cient\'{\i}ficas through the bilateral japanese-spanish project 2006JP0017.
Numerical computation in this work was carried out on Altix3700 BX2
at YITP in Kyoto University.


\begin{thebibliography}{99}

\bibitem{Halzen:2002pg}
F.~Halzen and D.~Hooper,
Rept.\ Prog.\ Phys.\  {\bf 65}, 1025 (2002)
[arXiv:astro-ph/0204527].

\bibitem{ICECUBE}
http://icecube.wisc.edu/

\bibitem{BAIKAL}
http://www.ifh.de/baikal/baikalhome.html

\bibitem{NESTOR}
http://www.nestor.org.gr/

\bibitem{ANTARES}
http://antares.in2p3.fr/index.html

\bibitem{NEMO}
http://nemoweb.lns.infn.it/project.htm

\bibitem{RICE}
http://www.bartol.udel.edu/~spiczak/rice/rice.html

\bibitem{AUGER}
http://www.auger.org/

\bibitem{EUSO}
http://euso.iasf-palermo.inaf.it/

\bibitem{Learned:1994wg}
J.~G.~Learned and S.~Pakvasa,
Astropart.\ Phys.\  {\bf 3}, 267 (1995)
[arXiv:hep-ph/9405296].

\bibitem{Bento:1999bb}
L.~Bento, P.~Keranen and J.~Maalampi,
Phys.\ Lett.\  B {\bf 476}, 205 (2000)
[arXiv:hep-ph/9912240].

\bibitem{Athar:2000yw}
H.~Athar, M.~Jezabek and O.~Yasuda,
Phys.\ Rev.\  D {\bf 62}, 103007 (2000)
[arXiv:hep-ph/0005104].

\bibitem{Barenboim:2003jm}
G.~Barenboim and C.~Quigg,
Phys.\ Rev.\  D {\bf 67}, 073024 (2003)
[arXiv:hep-ph/0301220].

\bibitem{Beacom:2003nh}
J.~F.~Beacom, N.~F.~Bell, D.~Hooper, S.~Pakvasa and T.~J.~Weiler,
Phys.\ Rev.\  D {\bf 68}, 093005 (2003)
[Erratum-ibid.\  D {\bf 72}, 019901 (2005)]
[arXiv:hep-ph/0307025].

\bibitem{Beacom:2002vi}
J.~F.~Beacom, N.~F.~Bell, D.~Hooper, S.~Pakvasa and T.~J.~Weiler,
Phys.\ Rev.\ Lett.\  {\bf 90}, 181301 (2003)
[arXiv:hep-ph/0211305].

\bibitem{Beacom:2003eu}
J.~F.~Beacom, N.~F.~Bell, D.~Hooper, J.~G.~Learned, S.~Pakvasa and T.~J.~Weiler,
Phys.\ Rev.\ Lett.\  {\bf 92}, 011101 (2004)
[arXiv:hep-ph/0307151].

\bibitem{Dutta:2001sf}
S.~I.~Dutta, M.~H.~Reno and I.~Sarcevic,
Phys.\ Rev.\  D {\bf 64}, 113015 (2001)
[arXiv:hep-ph/0104275].

\bibitem{Keranen:2003xd}
P.~Keranen, J.~Maalampi, M.~Myyrylainen and J.~Riittinen,
Phys.\ Lett.\  B {\bf 574}, 162 (2003)
[arXiv:hep-ph/0307041].

\bibitem{Awasthi:2007az}
R.~L.~Awasthi and S.~Choubey,
Phys.\ Rev.\  D {\bf 76}, 113002 (2007)
[arXiv:0706.0399 [hep-ph]].

\bibitem{Xing:2007rz}
Z.~z.~Xing,
arXiv:0711.4163 [astro-ph].

\bibitem{Pakvasa:2007dc}
S.~Pakvasa, W.~Rodejohann and T.~J.~Weiler,
JHEP {\bf 0802}, 005 (2008)
[arXiv:0711.4517 [hep-ph]].

\bibitem{Lipari:2007su}
P.~Lipari, M.~Lusignoli and D.~Meloni,
Phys.\ Rev.\  D {\bf 75}, 123005 (2007)
[arXiv:0704.0718 [astro-ph]].

\bibitem{Waxman:1997ti}
  E.~Waxman and J.~N.~Bahcall,
  Phys.\ Rev.\ Lett.\  {\bf 78}, 2292 (1997)
  [arXiv:astro-ph/9701231].
  
\bibitem{Fan:2008cg}
  Y.~Z.~Fan and T.~Piran,
  arXiv:0805.2221 [astro-ph].

\bibitem{Athanassopoulos:1996jb}
C.~Athanassopoulos {\it et al.}  [LSND Collaboration],
Phys.\ Rev.\ Lett.\  {\bf 77}, 3082 (1996)
[arXiv:nucl-ex/9605003].

\bibitem{Athanassopoulos:1997pv}
C.~Athanassopoulos {\it et al.}  [LSND Collaboration],
Phys.\ Rev.\ Lett.\  {\bf 81}, 1774 (1998)
[arXiv:nucl-ex/9709006].

\bibitem{Aguilar:2001ty}
A.~Aguilar {\it et al.}  [LSND Collaboration],
Phys.\ Rev.\  D {\bf 64}, 112007 (2001)
[arXiv:hep-ex/0104049].

\bibitem{Yao:2006px}
W.~M.~Yao {\it et al.}  [Particle Data Group],
J.\ Phys.\ G {\bf 33}, 1 (2006).

\bibitem{AguilarArevalo:2007it}
A.~A.~Aguilar-Arevalo {\it et al.}  [The MiniBooNE Collaboration],
Phys.\ Rev.\ Lett.\  {\bf 98}, 231801 (2007)
[arXiv:0704.1500 [hep-ex]].

\bibitem{Barger:2005mh}
V.~Barger, D.~Marfatia and K.~Whisnant,
Phys.\ Rev.\  D {\bf 73}, 013005 (2006)
[arXiv:hep-ph/0509163].

\bibitem{deGouvea:2006qd}
A.~de Gouvea and Y.~Grossman,
Phys.\ Rev.\  D {\bf 74}, 093008 (2006)
[arXiv:hep-ph/0602237].

\bibitem{Schwetz:2007cd}
T.~Schwetz,
JHEP {\bf 0802}, 011 (2008)
[arXiv:0710.2985 [hep-ph]].

\bibitem{Nelson:2007yq}
A.~E.~Nelson and J.~Walsh,
Phys.\ Rev.\  D {\bf 77}, 033001 (2008)
[arXiv:0711.1363 [hep-ph]].

\bibitem{Donini:2007yf}
A.~Donini, M.~Maltoni, D.~Meloni, P.~Migliozzi and F.~Terranova,
JHEP {\bf 0712}, 013 (2007)
[arXiv:0704.0388 [hep-ph]].

\bibitem{Maltoni:2004ei}
M.~Maltoni, T.~Schwetz, M.~A.~Tortola and J.~W.~F.~Valle,
New J.\ Phys.\  {\bf 6}, 122 (2004).
updated results in {\tt hep-ph/0405172 (v5)}.

\bibitem{Okada:1996kw}
N.~Okada and O.~Yasuda,
Int.\ J.\ Mod.\ Phys.\  A {\bf 12}, 3669 (1997)
[arXiv:hep-ph/9606411].

\bibitem{Bilenky:1996rw}
S.~M.~Bilenky, C.~Giunti and W.~Grimus,
Eur.\ Phys.\ J.\  C {\bf 1}, 247 (1998)
[arXiv:hep-ph/9607372].

\bibitem{Dydak:1983zq}
F.~Dydak {\it et al.},
Phys.\ Lett.\ B {\bf 134} (1984) 281.

\bibitem{Declais:1994su}
Y.~Declais {\it et al.},
Nucl.\ Phys.\ B {\bf 434} (1995) 503.

\bibitem{Barger:2000ch}
V.~D.~Barger, B.~Kayser, J.~Learned, T.~J.~Weiler and K.~Whisnant,
Phys.\ Lett.\  B {\bf 489}, 345 (2000)
[arXiv:hep-ph/0008019].

\bibitem{Sorel:2003hf}
M.~Sorel, J.~M.~Conrad and M.~Shaevitz,
Phys.\ Rev.\  D {\bf 70}, 073004 (2004)
[arXiv:hep-ph/0305255].

\bibitem{Karagiorgi:2006jf}
G.~Karagiorgi, A.~Aguilar-Arevalo, J.~M.~Conrad, M.~H.~Shaevitz, K.~Whisnant, M.~Sorel and V.~Barger,
Phys.\ Rev.\  D {\bf 75}, 013011 (2007)
[arXiv:hep-ph/0609177].

\bibitem{Maltoni:2007zf}
M.~Maltoni and T.~Schwetz,
Phys.\ Rev.\  D {\bf 76}, 093005 (2007)
[arXiv:0705.0107 [hep-ph]].

\bibitem{Bilenky:1998ne}
  S.~M.~Bilenky, C.~Giunti, W.~Grimus and T.~Schwetz,
  Astropart.\ Phys.\  {\bf 11}, 413 (1999)
  [arXiv:hep-ph/9804421].

\bibitem{Foot:1996qc}
  R.~Foot and R.~R.~Volkas,
  Phys.\ Rev.\  D {\bf 55}, 5147 (1997)
  [arXiv:hep-ph/9610229].

\bibitem{Cirelli:2004cz}
  M.~Cirelli, G.~Marandella, A.~Strumia and F.~Vissani,
  Nucl.\ Phys.\  B {\bf 708}, 215 (2005)
  [arXiv:hep-ph/0403158].

\bibitem{Dar:2003vf}
  A.~Dar and A.~De Rujula,
  Phys.\ Rept.\  {\bf 405} (2004) 203
  [arXiv:astro-ph/0308248].

\bibitem{Dar:2006dy}
  A.~Dar and A.~De Rujula,
  arXiv:hep-ph/0606199.

\bibitem{GonzalezGarcia:2005xw}
  M.~C.~Gonzalez-Garcia, F.~Halzen and M.~Maltoni,
  Phys.\ Rev.\  D {\bf 71} (2005) 093010
  [arXiv:hep-ph/0502223].

\bibitem{Lipari:2006uw}
  P.~Lipari,
  Nucl.\ Instrum.\ Meth.\  A {\bf 567} (2006) 405
  [arXiv:astro-ph/0605535].
  
\bibitem{Aharonian:2005jn}
  F.~Aharonian {\it et al.}  [The H.E.S.S. Collaboration],
  Science {\bf 307} (2005) 1938
  [arXiv:astro-ph/0504380].
  
\bibitem{Aharonian:2005kn}
  F.~Aharonian {\it et al.}  [HESS Collaboration],
  Astrophys.\ J.\  {\bf 636} (2006) 777
  [arXiv:astro-ph/0510397].
  
\bibitem{Kappes:2006fg}
  A.~Kappes, J.~Hinton, C.~Stegmann and F.~A.~Aharonian,
  Astrophys.\ J.\  {\bf 656} (2007) 870
  [arXiv:astro-ph/0607286].
  
\bibitem{Kistler:2006hp}
  M.~D.~Kistler and J.~F.~Beacom,
  Phys.\ Rev.\  D {\bf 74} (2006) 063007
  [arXiv:astro-ph/0607082].
 
\bibitem{EGRET}
EGRET Collaboration, Astrophys. \ J. {\bf 440} (1995) 525.

\bibitem{Maltoni:2008jr}
  M.~Maltoni and W.~Winter,
  arXiv:0803.2050 [hep-ph].
  
  


\end{thebibliography}
\end{document}